\documentclass[conference, 10pt]{IEEEtran}

\usepackage{caption}
\usepackage{xspace}
\usepackage{graphicx}
\usepackage[caption=false]{subfig}
\usepackage{amssymb}
\usepackage{pifont} 
\begin{document}

\newcommand\figref{Fig.~\ref}
\newcommand\eqnref{Equation~\ref}
\newcommand\secref{Section~\ref}
\newcommand\tabref{Table~\ref}
\newcommand\kstest{Kolmogorov–Smirnov test \xspace}
\newcommand\etal{et al.\xspace}

\newcommand\akash[1]{\says{Akash}{cyan}{#1}}
\newcommand\mani[1]{\says{Mani}{orange}{#1}}
\newcommand\luis[1]{\says{Luis}{pink}{#1}}
\newcommand\brian[1]{\says{Brian}{yellow}{#1}}
\newcommand\xac[1]{\says{Anthony}{green}{#1}}

\title{Understanding factors behind IoT privacy - A user's perspective on RF sensors}

\author{Akash Deep Singh$^\dagger$, Brian Wang$^\dagger$, Luis Antonio Garcia$^\S$, Xiang `Anthony' Chen$^\dagger$, Mani Srivastava$^\dagger$ \\
akashdeepsingh@g.ucla.edu, wangbri1@g.ucla.edu, la.garcia@utah.edu, xac@ucla.edu, mbs@ucla.edu \\
$^\dagger$University of California, Los Angeles (UCLA), $^\S$University of Utah}

\maketitle

\begin{abstract}
While IoT sensors in physical spaces have provided utility and comfort in our lives, their instrumentation in private and personal spaces has led to growing concerns regarding privacy. The existing notion behind IoT privacy is that the sensors whose data can easily be understood and interpreted by humans (such as cameras) are more privacy-invasive than sensors that are not human-understandable, such as RF (radio-frequency) sensors. However, given recent advancements in machine learning, we can not only make sensitive inferences on RF data but also translate between modalities. Thus, the existing notions of privacy for IoT sensors need to be revisited. In this paper, our goal is to understand what factors affect the privacy notions of a non-expert user (someone who is not well-versed in privacy concepts). To this regard, we conduct an online study of $162$ participants from the USA to find out what factors affect the privacy perception of a user regarding an RF-based device or a sensor. Our findings show that a user's perception of privacy not only depends upon the data collected by the sensor but also on the inferences that can be made on that data, familiarity with the device and its form factor as well as the control a user has over the device design and its data policies. When the data collected by the sensor is not human-interpretable, it is the inferences that can be made on the data and not the data itself that users care about when making informed decisions regarding device privacy. 
\end{abstract}

\section{Introduction}\label{sec:intro}

The ubiquitous growth of the Internet-of-Things (IoT) has resulted in sensors and sensing devices becoming an integral part of our daily lives – from smart homes and offices to public spaces. These sensors yield valuable data both for end-users (e.g., fitness activity tracking) and device owners (e.g., building security). However, concerns arise due to the opaque data handling by manufacturers and operators, sparking discussions on privacy, especially about sensor deployment and user disclosures~\cite{naeini2017privacy, igp2015protecting, emami-naeini_ask_2020}.

Discussions on sensor \textit{invasiveness} and \textit{sensitivity} typically revolve around safeguarding information reflecting users' actions and behaviors in the physical domain~\cite{lee2017privacy, psychoula2018users, gochoo_device-free_2018, zhou2015information}.

\noindent\textbf{The Challenge.} While privacy has multiple definitions (\secref{sec:terms}), recent IoT literature has adopted human-understandability as a measure of privacy. The prevailing view is that sensors capturing data discernible by human senses, like audio or visual recordings, should be absent from private zones such as bedrooms or restrooms~\cite{taylor2010spy, schwartz2012chicago}. This has spurred studies asserting that lower-dimensional RF sensors maintain better privacy compared to their visual counterparts~\cite{singh2019radhar, billah_ble_2021}.

\noindent\textbf{Underlying Assumption.} The tacit assumption in many works seems to be that non-expert users, unfamiliar with nuanced definitions of privacy, rely on their intuition, past experiences, and initial impressions to assess the privacy implications of a device. This paper delves into how such users perceive privacy when introduced to a sensor and given basic device information.

Though some studies posit that raw RF sensor data is less revealing than visual data~\cite{fan_-home_nodate}, it's worth noting that raw visual data isn't directly interpretable but undergoes transformation into recognizable images~\cite{isola2011understanding}. Analogously, low-dimensional RF sensor data can be processed to divulge sensitive information. This raises the question:
\newline 
\textit{\textbf{RQ1a}. Will translating non-human-understandable sensor data into a comprehensible format alter user privacy perceptions regarding that sensor?}

Considering humans may not decipher raw RF sensor data but can grasp algorithmic inferences from it, and given the significant strides in deep learning enabling various RF data inferences~\cite{zhao2016emotion, singh2019radhar, singh2023deep, singh2023depth}, it's pertinent to ask:
\newline 
\textit{\textbf{RQ1b}. When forming privacy opinions about a device, what weighs more for users: the data a device gathers or the possible inferences from that data, or both?}

Further, theories such as the mere-exposure effect suggest that object familiarity can influence human preferences~\cite{bornstein1992stimulus}. Thus, a related question emerges: 
\newline
\textit{\textbf{RQ2}. Does device familiarity sway a user's privacy perception?}

Moreover, as physical appearance can heavily impact first impressions and purchase decisions (as demonstrated by studies conducted on shoppers)~\cite{creusen2005different}, it's crucial to ascertain:
\newline
\textit{\textbf{RQ3}. How does a device's physical look impact user privacy perceptions?} 
For instance, would a benign-looking device, when designed like a camera, affect privacy perceptions differently?

Lastly, with rising concerns about data handling~\cite{wang2016big, wang2016using} and the unforeseen threats in digital devices~\cite{bilge2012before, rostami12being}, we explore:
\newline
\textit{\textbf{RQ4}. How do device design autonomy and data policies influence user privacy perceptions?} 

This research inspects user privacy perceptions surrounding low-dimensional sensor data, employing RF sensors like mmWave radar as exemplars, and endeavors to elucidate how factors like sensing modality, inferential capacity, familiarity, appearance, and control shape user perspectives.

\noindent\textbf{Study Details.} We executed an online survey on 162 US respondents to gauge privacy perceptions, focusing on data nature, device knowledge, appearance, and control. This study ascertains the role of each factor in sculpting user perceptions.

\noindent\textbf{Study Results.} Key insights include:
\begin{itemize}
    \item For human-interpretable data sensors (e.g., cameras), data and inferences guide user privacy decisions equivalently.
    \item For non-interpretable data sensors (e.g., RF sensors), inferences predominantly shape user privacy views.
    \item Absent data and inferences, a device's appearance and a user's familiarity predominantly steer privacy perceptions.
    \item Complete control over data policies and device design amplifies user willingness to integrate sensing devices in personal spaces.
\end{itemize}

\noindent\textbf{Research Aims.} We strive to discern factors influencing non-expert user privacy perceptions upon encountering an IoT device, and aspire to enrich future sensor deployments with our insights.

\noindent\textbf{Contributions.} The main contributions offered by our paper are:
\begin{itemize}
    \item A robust exploration of factors determining non-expert privacy perspectives in light of deep learning advances for low-dimensional sensor data – a pioneering effort to our knowledge.
    \item A user study elucidating how these factors mold user privacy perceptions.
    \item Recommendations for researchers to amplify privacy disclosures across sensing domains.
\end{itemize}

\section{Related Work}\label{sec:related_work}

This section reviews research on RF's privacy sensitivity compared to visual modalities, the influence of sensor characteristics on privacy perceptions, and efforts to understand and measure privacy behavior of sensors and systems. To our knowledge, our study is the first to juxtapose device-collected data with inferred information and its impact on privacy perception. We explored factors such as physical appearance, familiarity, device type, inferences, data modality, and user control over data policies.

\subsection{Claims without validation: RF versus Camera/Vision}

A significant trend in the RF domain suggests using RF sensing as a less privacy-intrusive alternative to cameras. These claims arise from studies spanning various applications, from monitoring building occupancy~\cite{billah_ble_2021, aziz_shah_privacy-preserving_2020} to detecting breathing patterns~\cite{ashleibta_non-invasive_2021}. A common thread in these works is the unvalidated claim of RF's inherently superior privacy attributes over cameras. Our research challenges these assertions, highlighting unforeseen privacy-invasive inferences across modalities and the need for a comprehensive understanding of what defines a modality as "privacy invasive".

\subsection{Exploring Privacy Sensitivity of Devices and Sensors}

The perception of a sensor's privacy is multifaceted, influenced by its design, deployment context, and the reputation of its manufacturer. Studies have delved into the role of physical design in privacy perception, with concepts like 'tangible privacy'~\cite{ahmad_tangible_2020} or designs clearly indicating sensing state~\cite{koelle_beyond_2018}. Manufacturer trust is also paramount; brand recognition and reputation significantly influence trust and purchasing decisions~\cite{zheng_user_2018, malkin2019privacy}. Apart from these, other works have probed the influence of data-sharing behaviors and user familiarity with the technology on privacy perceptions~\cite{naeini2017privacy, lee2022understanding}. Our study, focusing on RF and camera modalities, seeks to contextualize these insights to understand the variables shaping privacy perceptions.

\subsection{Structuring and Measuring Privacy Properties}

Research has emphasized creating transparency mechanisms to elucidate privacy properties of devices and systems. Companies like Apple~\cite{noauthor_privacy_nodate} and Google~\cite{noauthor_get_2022} have initiated efforts to clarify app installations' privacy implications. However, potential privacy-invasive inferences often remain unaddressed. Alongside, various works aim to quantify the privacy attributes of systems, whether through grading mobile applications~\cite{noauthor_privacygrade_nodate} or ranking websites based on security features~\cite{noauthor_welcome_nodate}. Distinct from these, our study is anchored in the human perspective, aiming to redefine privacy conceptions surrounding RF sensing.

\section{Privacy Conceptions and Terms in this Paper}\label{sec:terms}

In this paper, we use the terms privacy notion, privacy perception, and privacy preference interchangeably to describe how a user perceives a device or a sensor from a privacy perspective - does the user think a device is going to protect their privacy or not? A device that is designed with privacy in mind (or is perceived to protect user privacy) is considered less privacy-invasive and more privacy-sensitive.

\subsection{Popular Conceptions of Privacy}

There are several conceptions of privacy~\cite{nissim2018privacy, solove2008understanding, nissenbaum2009privacy, cranor2013conceptions}. In this section we discuss some popular conceptions. The definitions for these conceptions have been borrowed from the works cited above.

\begin{itemize}
    \item \textbf{Contextual Integrity}: Contextual integrity assesses how closely the flow of personal information conforms to context-relative informational norms. More precisely, in a context, the flow of information of a certain type about a subject from a sender to a recipient is governed by a particular transmission principle. Contextual integrity is violated when the norms in the relevant context are breached. Intuitively, it recognizes that certain parties may obtain certain types of information about other parties under the right terms and for the right reasons.
    \item \textbf{Anonymization and De-identification}: Many privacy technologies are designed with the goal of de-identifying personal information. This approach equates privacy protection with making personal information anonymous or de-identified, i.e. preventing an individual’s information from being linked with them. The premise is that it is impossible (or, at least, very difficult) to infer personal information pertaining to an individual from a de-identified dataset or use it to violate an individual’s privacy in other ways.
    \item \textbf{Semantic Security}: The definition of semantic security compares what an attacker (without access to the decryption key) can predict about the message m given the ciphertext c with what the attacker can predict about the message m without being given the ciphertext c. The advantage that access to the ciphertext gives to any attacker is quantified. Encryption schemes are designed to make this advantage so negligible that access to the ciphertext does not give the attacker any practical advantage over not getting any information about the message at all.
    \item \textbf{Differential Privacy}: It guarantees mathematically that a person, who is observing the outcome of a differential private analysis, will produce likely the same inference about an individual’s private information, whether or not that individual’s private information is combined in input for the analysis.
\end{itemize}

We suspect that the IoT sensing papers that use human-understandability as a proxy for privacy assume that when encountering a new sensor, a non-expert user will have no knowledge of these conceptions and hence will rely more on their experiences with the device or the class of devices. The privacy notions formed during this interaction between a non-expert user and a sensor is what we aim to study in this paper. Additionally, we also believe that the factors discussed here can be used in conjunction with the above definitions to better inform the design for privacy-centric sensor installations.
\section{Method}\label{sec:method}
This section details our survey and data analysis methodology.

\subsection{Hypothesis}
We aim to understand factors influencing users' perceptions of device privacy. Four primary factors are identified:

\subsubsection{Privacy Factor (1): User's Prior Knowledge}
Leveraging the mere-exposure effect~\cite{bornstein1992stimulus, bornstein2016mere}, we hypothesize that familiarity with a device or its properties influences its perceived privacy: \textit{A user's familiarity with a device, including its manufacturer and type, affects their privacy perceptions about that device.}

\subsubsection{Privacy Factor (2): Device Appearance}
Considering that visual impressions impact perceptions, we hypothesize that altering a device's appearance impacts its perceived privacy: \textit{Changing a device's appearance, even if the user remains familiar with it, will lead to more negative privacy perceptions.}

\subsubsection{Privacy Factor (3): Data Interpretability}
Given that less-interpretable data is deemed more private~\cite{li2019survey, fan_-home_nodate, raeis_human_2021, avrahami_below_2018}, we aim to evaluate whether user concerns lie more with the type of data collected or its potential inferences: \textit{Awareness of inferences from sensor data may alter a user's privacy perception about the device.}

\subsubsection{Privacy Factor (4): Control over Data}
In an era of data-driven decisions~\cite{wang2016big, hochheiser2015truth, wang2016using}, control over data affects user willingness to adopt devices. We hypothesize that: \textit{Greater control over data collection and storage increases users' willingness to use sensing devices.}

\subsection{Survey Study}
Our online survey, taking 12-18 minutes, gauged users' privacy preferences. We presented participants with images of sensing devices, data they collect, and potential inferences, alongside situational questions to assess our hypothesis.

\subsubsection{Scenario Selection}
To understand the primary privacy concern among physical appearance, data representation, or data inferences, we employed a three-step approach: 
1) Introduce a device through an image.
2) Ask users for their familiarity and perception of the device.
3) Assess comfort with the device in their private spaces.
The study compared perceptions of cameras, mmWave sensors, WiFi routers, and audio sensors, considering various data control scenarios.

\begin{figure*}[!tbp]
    \centering
    \subfloat[]{\includegraphics[width=0.18\textwidth]{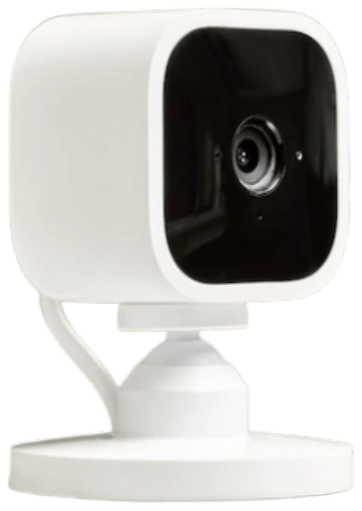}}
    \subfloat[]{\includegraphics[width=0.1\textwidth]{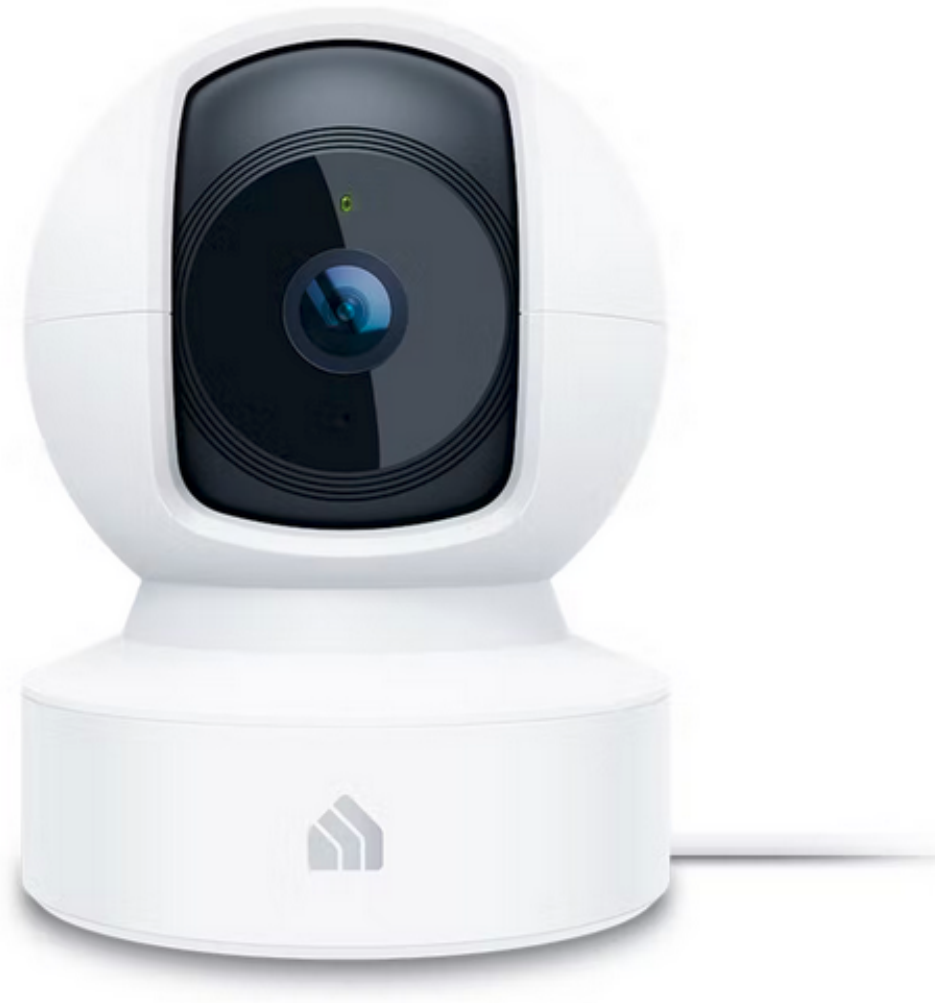}}
    \subfloat[]{\includegraphics[width=0.18\textwidth]{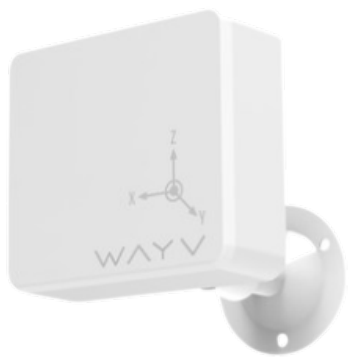}}
    \subfloat[]{\includegraphics[width=0.15\textwidth]{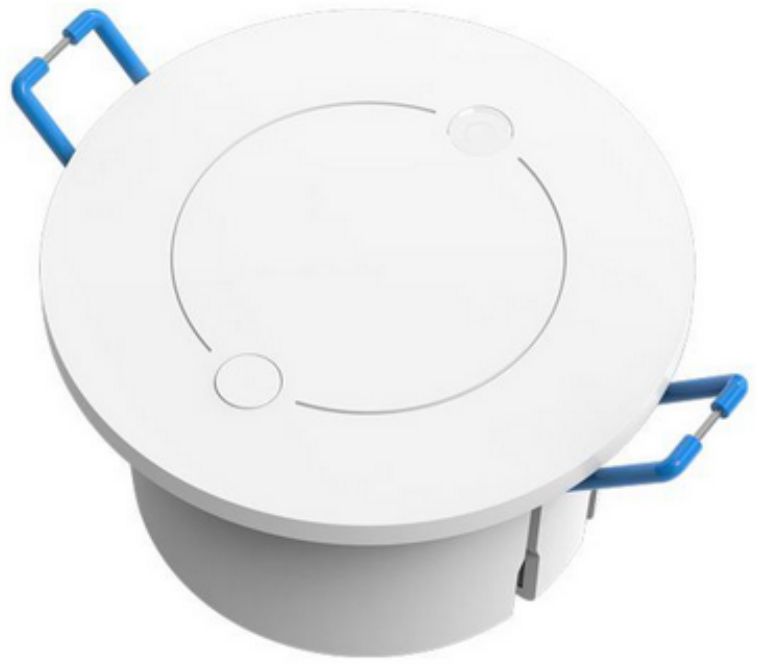}}
    \subfloat[]{\includegraphics[width=0.2\textwidth]{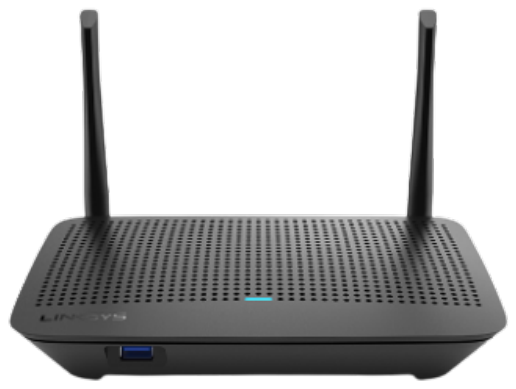}}
    \subfloat[]{\includegraphics[width=0.15\textwidth]{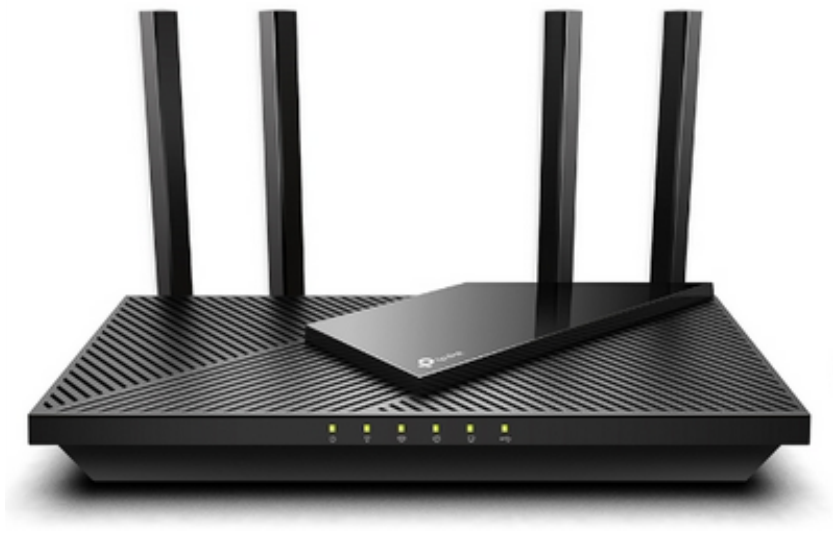}}
    
    \caption{We use 3 main classes of devices in our study. Each class has 2 different devices (in order to ascertain that the user responses are dependent on the class of device and not on the device itself.) These devices are \textbf{a)} An Amazon Blink home camera, \textbf{b)} Kasa Inoor Pan/Tilt Smart Security Camera \textbf{c)} A Wayv mmWave radar from Aienstein AI, \textbf{d)} LifeSmart mmWave Human Presence Sensor, \textbf{e)} LinkSys MAX-STREAM AC1300 WiFi router, and \textbf{f)} TP-Link AX1800 WiFi 6 router. In this paper, we use RF devices to learn more about factors that govern user privacy perceptions.}
    \label{fig:survey_devices_methods}
\end{figure*}

\subsubsection{Device Selection}
Our survey featured three device classes: cameras, RF sensors, and WiFi routers. To ensure the device class, not the brand or model, influenced responses, we showcased two sets of images for each device type (as depicted in \figref{fig:survey_devices_methods}). Devices were selected based on their commercial availability and popularity.

\subsubsection{Sample Data Representations for Devices}
For each device, we provided users with representative data samples:
\begin{itemize}
    \item \textbf{Camera:} Image of two individuals sleeping.
    \item \textbf{mmWave Radar:} Snapshot od radar output from the company website.
    \item \textbf{WiFi Router:} Wireshark packet sniffing output~\cite{banerjee2010evaluation}.
\end{itemize}

\subsubsection{Device Inferences}
Respondents were shown potential inferences derived from the data collected by each device, which are listed below:
\begin{itemize}
    \item \textbf{Camera:} (i) The number of people in the room ~\cite{teixeira2007lightweight}, their clothing~\cite{stearns2018applying, huang2021clothing}, race~\cite{layne2012person}, ethnicity~\cite{layne2012person}, body shape~\cite{guan2009estimating}, height~\cite{guan2009estimating}, and posture~\cite{guan2009estimating}; (ii) Certain activities, behaviors, and health conditions of the people~\cite{ann2014human, demrozi2020human}: having meals, drinking, smoking, walking, praying, watching TV, using a smartphone, intimate moments (hugging, kissing), and breathing rate~\cite{massaroni2018contactless}.
    \item \textbf{mmWave Radar:} (i) The number of people in the room ~\cite{choi2017people, weiss2020improved}, body shape, height, and posture~\cite{li2020capturing, sengupta2020mm}; (ii) Certain activities, behaviors, and health conditions of the people~\cite{li2019survey, ding2021radar, singh2019radhar}: having meals, drinking, smoking, walking, praying, watching TV, using a smartphone, intimate moments (hugging, kissing), and breathing rate~\cite{wang2020remote, chauhan2020through}.
    \item \textbf{WiFi Router:} (i) Websites that have been visited, total time spent on a smartphone, and sleep schedule~\cite{ohm2009rise, hernandez2021reduction}; (ii) The number of people in the room ~\cite{yang2018wi, cheng2017device}, body shape, height, and posture~\cite{jiang2020towards, ren2022gopose}; (iii) Certain activities, behaviors, and health conditions of the people~\cite{wang2017device, li2019wi, chen2018wifi}: having meals, walking, praying, watching TV, using a smartphone, intimate moments (hugging, kissing), and breathing rate~\cite{abdelnasser2015ubibreathe, gao2020device}.
\end{itemize}

\subsubsection{Privacy Perception Metrics}\label{sec:data_inferences_privacy}
We assessed users’ privacy concerns using three questions:
\begin{enumerate}
    \item Perception based on the device's appearance.
    \item Comfort level considering the type of data the device collects.
    \item Comfort level with device inferences, ensuring data isn't viewed by humans.
\end{enumerate}

\subsubsection{Relating Privacy and Comfort}
We equated "privacy-sensitivity" and "privacy-invasiveness" with user comfort for clarity, in line with existing studies~\cite{akter_i_2020}.

\subsubsection{Response Scale}
Respondents used a 5-point Likert scale, with categories for familiarity and comfort.

\subsection{Survey Setting}
Considering bedrooms as the most private spaces~\cite{berry_2020}, we used them as a context to understand users' privacy perceptions.

\subsection{Survey Structure}
The survey comprised:
\begin{itemize}
    \item Consent and demographic forms.
    \item Sections evaluating comfort with devices in bedrooms based on appearance, data, and inferences.
    \item Scenario of a self-made camera.
\end{itemize}

\subsection{Data Analysis}
\subsubsection{Quantitative Analysis}
Given non-normal distribution of responses, non-parametric tests (Wilcoxon signed-rank, \kstest, Spearman correlation) were utilized. Summary statistics were also computed.

\subsubsection{Qualitative Analysis}
Open-ended responses were analyzed using inductive coding, with inter-researcher agreement assessed via Cohen's Kappa.

\subsection{Recruitment}
The study was conducted on Prolific~\cite{palan2018prolific} in mid-2022, targeting US participants aged 18+.

\subsection{Ethical Aspects}
Participants, recruited via Prolific, could opt out after viewing the Google form consent. Participation was voluntary with transparency on data usage.

\section{Findings: Quantitative Analysis}\label{sec:quant_find}

In this section, we describe the quantitative findings of our survey.

\subsection{Age and Gender}
A total of 162 respondents completed the survey on Prolific. Out of the 162 participants, 94 (58\%) identified as male, 64 (39.5\%) identified as female, 3 (1.85\%) neither identified as male nor as female and 1 chose to not disclose their gender.

Amongst the 162 participants, 5 (3.1\%) were less than 20 years of age (and 18 years or older), 110 (67.9\%) were between 21 and 40 years of age, 39 (24.1\%) were between 41 and 60 years of age and 8 (4.9\%) were above 60 years of age.

\subsection{Education and Technical Level}
During the survey, we asked the respondents to self-identify their education level. We asked them to select the highest level of education that they have achieved. 52 (32.1\%) of the total respondents' reported their highest level of education as high-school, 23 (14.2\%) as Associates, 51 (31.5\%) as Bachelors, 32 (19.8\%) as Graduate and 4 (2.5\%) as Professional.

In order to assess the technical savviness and familiarity of the respondents with sensing devices and the internet-of-things (IoT) in general, we used modified versions of some questions from the Mozilla's `How connected are you?' survey~\cite{mozilla}. We asked the respondents: \textit{How would you describe yourself when it comes to your knowledge of information technology?} The options were:
\begin{itemize}
    \item \textit{I am an expert: I build my own technical systems (e.g., computers), run my own servers, and code my own apps.}
    \item \textit{I am technically savvy: I know my way around a computer pretty well. When anyone in my family needs technical help, I’m the one they call.}
    \item \textit{I am an average user: I know enough to get by.}
    \item \textit{I am a novice: Technology scares me! I only use it when I have to.}
\end{itemize}

Out of the total 162 respondents, 17 (10.5\%) considered themselves experts, 94 (58\%) considered themselves technically-savvy, 48 (29.6\%) considered themselves average and 3 (1.9\%) considered themselves as novice.

\subsection{Data vs Inferences}

In this section, we investigate user perceptions of device privacy based on data interpretability, possible inferences, or a combination of both. Using three questions detailed in \secref{sec:data_inferences_privacy}, we aim to distinguish between the impacts of data and inferences on privacy concerns. Initially, users view only the device image to gauge comfortability with its presence in their bedrooms. This assesses any biases based on device appearance. Next, alongside the device image, we present a data snapshot collected by the device, examining the role of data interpretability in shaping privacy perceptions. In the final question, users are presented with potential inferences from the data, assuring them that only algorithms will process the data and only inferences will be shared. This explores the influence of knowing inferences on privacy views. Quantitative results for these questions across different sensing devices are presented in \figref{fig:survey_devices_methods}.

\subsubsection{Camera}

\begin{figure*}[h]
\centering
\includegraphics[width=0.65\linewidth]{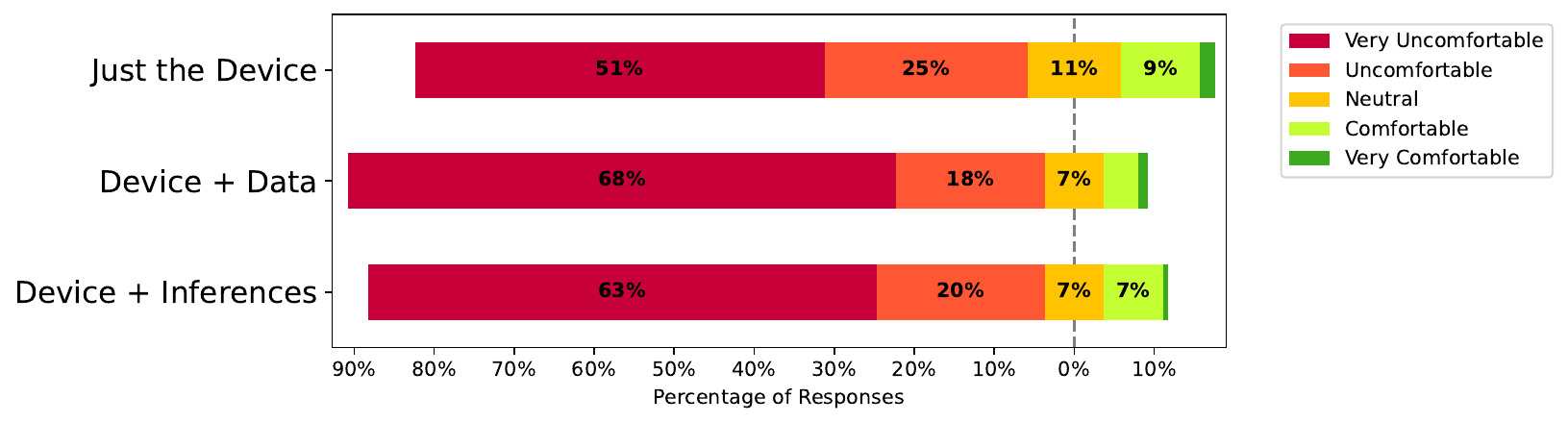}
\caption{User comfort level when shown (a) an image of the camera, (b) an image of the camera and a snapshot of the data that it collects, (c) an image of the camera and list of inferences that can be made on the data that it collects (assuming that only inferences and not data are being shared)}
\label{fig:camera_responses}
\end{figure*}

We showed the users pictures of a camera (\figref{fig:survey_devices_methods}) and asked about their comfort level if it were to be placed in their bedroom. Since cameras are ubiquitous today, only 22 out of the 162 respondents said that they were either unfamiliar or very unfamiliar with it. Additionally, since cameras are considered to be privacy invasive, we found that people were not comfortable with installing it in their bedrooms~\figref{fig:camera_responses}. 124 out of the 162 total respondents ($\mu = 1.85$, $\sigma = 1.08$, 95\% CI = [1.69, 2.03]) were either very uncomfortable or uncomfortable when just shown the image of the camera. Next, we showed the respondents an image of the same camera, but with a snapshot of a sample data (image) captured by it. We asked the users, the same question regarding their comfort level with installing this device in their bedrooms. 141 out of the 162 total respondents ($\mu = 1.51$, $\sigma = 0.90$, 95\% CI = [1.37, 1.65]) were either very uncomfortable or uncomfortable, as shown in ~\figref{fig:camera_responses}. We performed a Wilcoxon signed-rank test between the responses after looking at just the physical appearance and the responses after looking at both the physical appearance and the data collected by the camera and found that the difference was statistically significant ($V=306.5$, $p < 0.0000054$) meaning that showing the collected data representation from the sensor caused a significant shift in their privacy perception. Next, we showed the users an image of the same camera with a list of inferences that can be made on the data collected by it (with the condition that humans won't be able to see the raw data). 137 out of the 162 total respondents ($\mu = 1.60$, $\sigma = 0.95$, 95\% CI = [1.46, 1.75]) were either very uncomfortable or uncomfortable, as shown in ~\figref{fig:camera_responses}. We performed a Wilcoxon signed-rank test between the responses after looking at just the physical appearance and the responses after looking at both the physical appearance and the list of inferences that can be made by the camera and found that the difference was statistically significant ($V=700.5$, $p < 0.00082$). Additionally, we did not find a statistically significant difference between the responses after looking at the physical appearance of the camera with the data that it collects and the responses after looking at both the physical appearance and the list of inferences that can be made by the camera ($V=270.5$, $p = 0.07$). To further analyze the relationship between the data and the inferences, we perform a \kstest between the two and find that the two distributions are similar ($D=0.049$, $p = 0.9895$). This means that \textbf{for a camera, both the data representations and the inferences lead to similar privacy perceptions.} This can be attributed to the fact that since users can look at camera images and make their own inferences, both data and inferences contains similar amounts of human understandable information.

\subsubsection{mmWave Radar}

\begin{figure*}[h]
\centering
\includegraphics[width=0.65\linewidth]{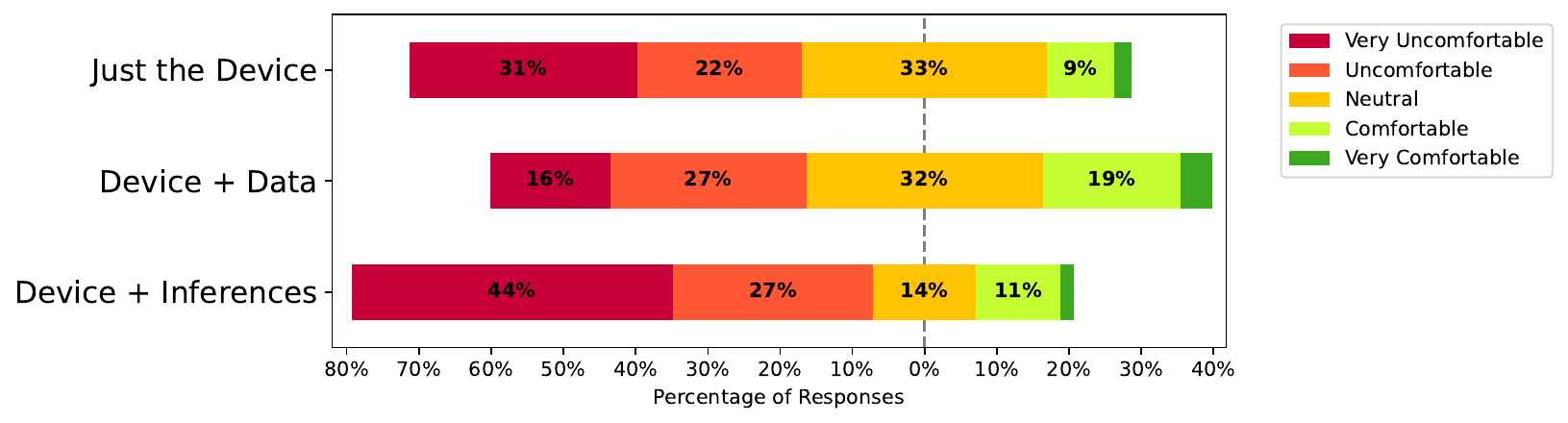}
\caption{User comfort level when shown (a) an image of the mmWave radar, (b) an image of the mmWave radar and a snapshot of the data that it collects, (c) an image of the mmWave radar and list of inferences that can be made on the data that it collects (assuming that only inferences, not data, are being shared)}
\label{fig:radar_responses}
\end{figure*}

We showed the users picture of an off-the-shelf mmWave radar (\figref{fig:survey_devices_methods}) and asked about their comfort level with placing it in their bedroom. mmWave radars are relatively new in the sensing world and hence, only 3 out of the 162 respondents said that they were familiar with it. We found that when shown an image of just the radar, users were not comfortable with installing it in their bedrooms~\figref{fig:radar_responses}. 88 out of the 162 total respondents ($\mu = 2.28$, $\sigma = 1.08$, 95\% CI = [2.12, 2.45]) were either very uncomfortable or uncomfortable. Next, we showed the respondents an image of the same radar but with a snapshot of a sample data captured by it. We asked the users the same question regarding their comfort level with installing this device in their bedrooms. Since the data captured by the radar is not human interpretable, the number of users that were either very uncomfortable or uncomfortable decreased to only 71 out of the 162 total respondents (shown in ~\figref{fig:radar_responses}) ($\mu = 2.67$, $\sigma = 1.09$, 95\% CI = [2.50, 2.84]). We performed a Wilcoxon signed-rank test between the responses after looking at just the physical appearance and the responses after looking at both the physical appearance and the data collected by the mmWave radar and found that the difference was statistically significant ($V=1018.5$, $p < 2.95\times 10^{-5}$) meaning that showing users the data collected by the sensor caused a significant shift in their privacy perception. Next, we showed the users an image of the same mmWave radar with a list of inferences that can be made on the data collected by it (with the condition that humans won't be able to see the raw data). 117 out of the 162 total respondents ($\mu = 1.99$, $\sigma = 1.11$, 95\% CI = [1.82, 2.16]) were either very uncomfortable or uncomfortable (shown in ~\figref{fig:radar_responses}). We performed a Wilcoxon signed-rank test between the responses after looking at just the physical appearance and the responses after looking at both the physical appearance and the list of inferences that can be made by the mmWave radar and found that the difference was statistically significant ($V=1589.5$, $p = 0.0023$). We also found a statistically significant difference between the responses after looking at the physical appearance of the mmWave radar with the data that it collects and the responses after looking at both the physical appearance and the list of inferences that can be made by the mmWave radar ($V=596.0$, $p < 4.86 \times 10^{-11}$). To further analyze the relationship between the data and the inferences, we perform a \kstest between the two and find that their is no similarity between the two distributions ($D=0.284$, $p < 3.70 \times 10^{-6}$). We conclude that \textbf{for a mmWave radar, since the data is not human understandable but inferences are, the data and the inferences lead to different privacy notions. In fact, when shown the inferences, the users' privacy perceptions become more negative.} This shows that inferences lead to more informed privacy decisions by the users since mmWave radar data is not human interpretable, whereas inferences are. Additionally, since people are not familiar with mmWave sensors, their unfamiliarity initially dominates their decision making about placing this sensors in their bedrooms.

\subsubsection{WiFi Router}

\begin{figure*}[h]
\centering
\includegraphics[width=0.65\linewidth]{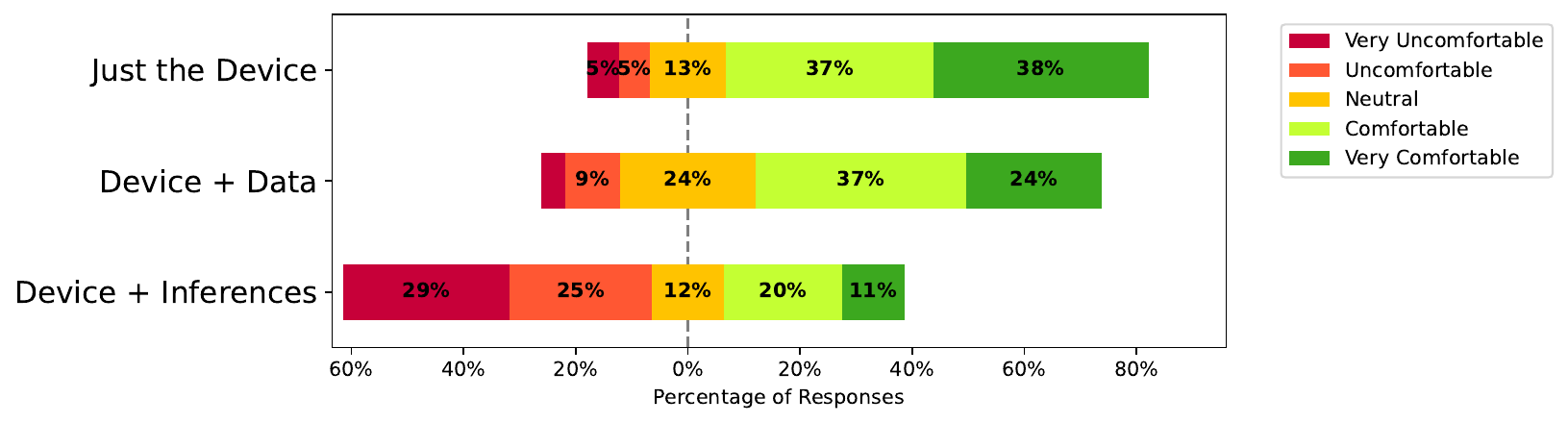}
\caption{User comfort level when shown (a) an image of the WiFi router, (b) an image of the WiFi router and a snapshot of the data that it collects, (c) an image of the WiFi router and list of inferences that can be made on the data that it collects (assuming that only inferences and not data are being shared)}
\label{fig:router_responses}
\end{figure*}

We showed the users pictures of a generic WiFi router (\figref{fig:survey_devices_methods}) and asked about their comfort level with placing it in their bedroom. Since WiFi routers are ubiquitous today, only 8 out of the 162 respondents said that they were either unfamiliar or very unfamiliar with it. Additionally, since WiFi routers are considered to be innocuous, we found that people were very comfortable with installing it in their bedrooms (\figref{fig:router_responses}). 122 out of the 162 total respondents ($\mu = 3.97$, $\sigma = 1.11$, 95\% CI = [3.80, 4.14]) were either very comfortable or comfortable when just shown the image of the WiFi router. Next, we showed the respondents an image of the same WiFi router but with a snapshot of a sample data (Wireshark snapshot) captured by it. We asked the users the same question regarding their comfort level with installing this device in their bedrooms. Since the Wireshark output is not interpretable by regular users, 100 out of the 162 total respondents were either very comfortable or comfortable (\figref{fig:router_responses}) ($\mu = 3.67$, $\sigma = 1.08$, 95\% CI = [3.51, 3.84]). We performed a Wilcoxon signed-rank test between the responses after looking at just the physical appearance and the responses after looking at both the physical appearance and the data collected by the WiFi router and found that the difference was statistically significant ($V=647.0$, $p < 3.83 \times 10^{-5}$) meaning that showing users the data collected by the sensor caused a significant shift in their privacy perception. Next, we showed the users an image of the same WiFi router with a list of inferences that can be made on the data collected by it (with the condition that humans won't be able to see the raw data). 89 out of the 162 total respondents were now either very uncomfortable or uncomfortable~\figref{fig:router_responses} ($\mu = 2.59$, $\sigma = 1.39$, 95\% CI = [2.37, 2.80]). The total number of uncomfortable users went from 13 (when shown just the physical appearance of the WiFi router) initially to 61 (when also shown the list of inferences). We performed a Wilcoxon signed-rank test between the responses after looking at just the physical appearance and the responses after looking at both the physical appearance and the list of inferences that can be made by the WiFi router and found that the difference was statistically significant ($V=271.0$, $p = 8.63 \times 10^{-19}$). We also found a statistically significant difference between the responses after looking at the physical appearance of the WiFi router with the data that it collects and the responses after looking at both the physical appearance and the list of inferences that can be made by the WiFi router ($V=237.5$, $p < 2.26 \times 10^{-16}$). To further analyze the relationship between the data and the inferences, we perform a \kstest between the two and find no similarity between the two ($D=0.410$, $p < 2.07 \times 10^{-12}$). We conclude that \textbf{showing inferences for a common home device like a WiFi router causes a very large negative shift in the privacy perception of that device.} This shift in privacy perception is different from the one which happens when users are shown the data representation, since the data collected is not human interpretable. Hence, \textbf{for sensors that collect non-human-interpretable data, it is the inferences, and not the data collected, that have a higher impact on how users perceive the privacy-sensitivity of a device.}

\subsection{Physical Appearance of the Device and Familiarity}
\begin{figure}[h]
\centering
\includegraphics[width=0.96\linewidth]{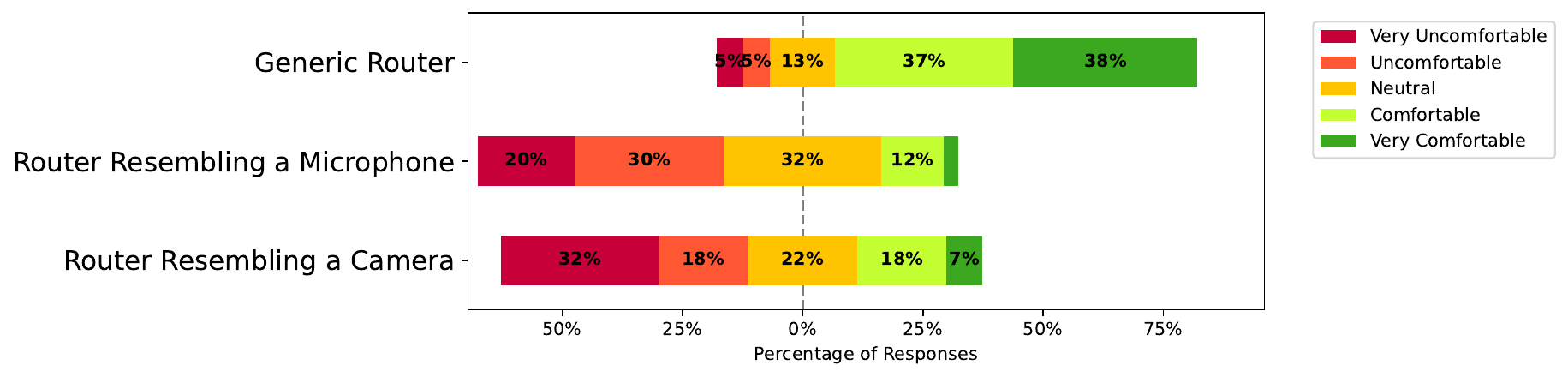}
\caption{User comfort level when shown (a) an image of a generic WiFi router, (b) an image of a WiFi router that looks like a microphone, (c) an image of a WiFi router that looks like a camera.}
\label{fig:router_change_in_appearance}
\end{figure}

WiFi routers are one of the most common IoT devices in homes in the US~\cite{parksassociates} -- something that is also reflected in the responses to the question -- \textit{How familiar are you with the following device/sensor} [with an image of a generic WiFi router]. Hence, it is not surprising to see that most of the users thought of it as a benign device and were comfortable with installing it in their bedrooms. In order to understand how physical appearance affected perceptions of privacy, we found two WiFi router designs that look drastically different. One is an Asus Blue Cave Wifi router that looks like a camera from certain angles, and second is a Maurice Misho Radar Router Design that looks like a microphone. The responses are shown in~\figref{fig:router_change_in_appearance}. Initially, when shown the image of a generic WiFi router, only 18 ($11.1\%$) of the total respondents ($\mu = 3.97$, $\sigma = 1.11$, 95\% CI = [3.80, 4.14]) were uncomfortable with installing it in their bedrooms. However, when shown a WiFi router that looks like a microphone, 83 ($51.2\%$) ($\mu = 2.48$, $\sigma = 1.05$, 95\% CI = [2.31, 2.64]) were uncomfortable and when shown a WiFi router that looks like a camera, 83 ($51.2\%$) out of the total 162 ($\mu = 2.49$, $\sigma = 1.31$, 95\% CI = [2.29, 2.70]) respondents were uncomfortable. We performed a Wilcoxon signed-rank test between responses for a generic WiFi router and a WiFi router that looks like a microphone and found that the difference was statistically significant ($V=140.5$, $p < 1.981 \times 10^{-21}$). Similarly, the difference between responses for a generic WiFi router and a WiFi router that looks like a camera was also statistically significant ($V=227.5$, $p < 1.27 \times 10^{-18}$). We also performed a Wilcoxon signed-rank test between responses for a WiFi router that looks like a microphone and a WiFi router that looks like a camera but did not find a statistically significant difference ($V=1629.5$, $p = 0.73$). Additionally, to further analyze the relationship between the responses for a WiFi router that looks like a microphone and a WiFi router that looks like a camera, we performed a Spearman Correlation Coefficient Test between the two and found a moderate Spearman correlation ($R=0.47$, $p < 4.385 \times 10^{-10}$). Hence, we conclude that \textbf{when the data collected and inferences are not disclosed to users, the physical appearance of a device can be manipulated to influence the users' privacy perception of that device.}

\subsection{Control}

\begin{figure}[h]
\centering
\includegraphics[width=0.99\linewidth]{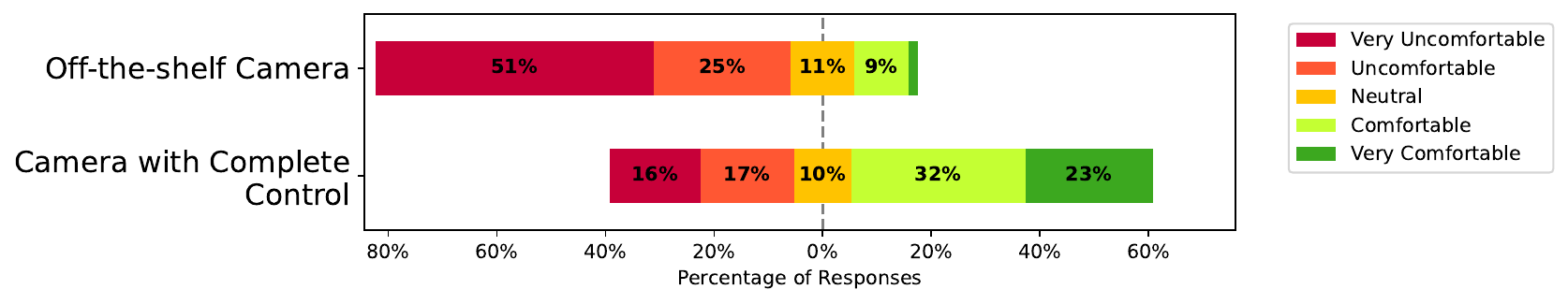}
\caption{User comfort levels with installing a camera in their bedroom when given full control over it's design and data policies. Notice that more users are willing to put a camera in their bedrooms if given more control.}
\label{fig:camera_control}
\end{figure}

To find out how control over device configuration and data policies affects a user's perception on privacy, we presented the respondents with a scenario where they get to build a camera from scratch -- they can control the physical appearance, hardware components, data collected and how the data is stored. We then asked the same question regarding their comfort levels with placing this camera in their bedrooms. The user responses are shown in~\figref{fig:camera_control}. We see that initially, only 19 ($11.7\%$) ($\mu = 1.85$, $\sigma = 1.08$, 95\% CI = [1.69, 2.03]) out of the total 162 respondents were comfortable with placing a third party camera in their bedrooms, however, when given complete control over the device, 90 ($55.6\%$) out of the total 162 respondents ($\mu = 3.28$, $\sigma = 1.42$, 95\% CI = [3.06, 3.51]) are comfortable with placing this camera in their bedrooms. We performed a Wilcoxon signed-rank test between two sets of responses and found that the difference was statistically significant ($V=284.0$, $p < 1.26 \times 10^{-18}$). Additionally, we also performed the Kolmogorov-Smirnov test between the two sets of responses and found that there indeed was a statistically significant difference between the two ($D=0.438$, $p < 2.31 \times 10^{-14}$). Hence, we conclude \textbf{when given complete control over the device design and data policies, users are more likely to instrument their personal spaces with sensing devices.}

\section{Findings: Qualitative Analysis}\label{sec:qual_find}

We present an analysis of user responses to a survey question probing their comfort level with a self-made camera in their bedroom, and the control over its design and data:

\noindent\textit{OE1. Suppose you create your own camera from scratch. You control what the camera looks like, what parts go inside it, what data it collects, and where the data is stored. How comfortable would you be with that camera being placed in your bedroom given that no one else can see the data except you? Please explain the choice you made in the previous question briefly.}

Responses were grouped by sentiment, coded as either positive (comfortable with the camera) or negative (uncomfortable). The analysis was performed by two researchers, using open-ended answers. 

\begin{figure}[h]
\centering
\includegraphics[width=0.96\linewidth]{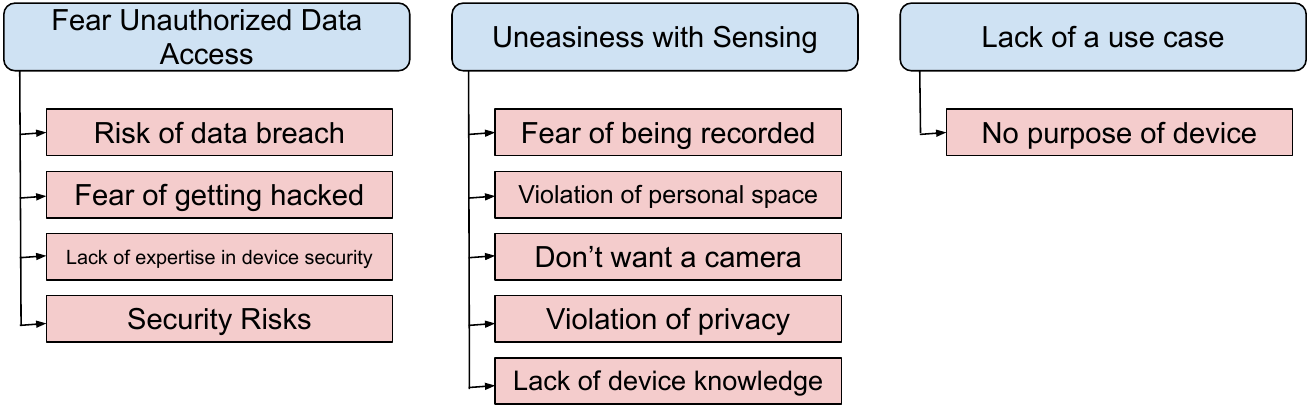}
\caption{Inductive coding of negative sentiment responses: themes (blue) and codes (red).}
\label{fig:code_neg}
\end{figure}

\subsection{Reasons Against Camera Placement}

Negative responses led to ten codes, organized into three themes, as seen in \figref{fig:code_neg}.

\subsubsection{Unauthorized Data Access}
Of 162 respondents, 43 (26.5\%) voiced concerns about unauthorized access, even for self-made devices. Concerns ranged from data breaches to hacking. One noted, \textit{"All systems are hackable... Privacy does not exist in this world."} Despite full control, many still felt uneasy about data vulnerability, with some more wary of their own creations than manufacturer-made devices.

\subsubsection{Discomfort with Sensing}
30 respondents (18.5\%) resisted cameras in their bedrooms, citing feelings of surveillance or intrusion. Comments included \textit{"It would still feel like there is something watching me..."} and \textit{"I don't like cameras in my bedroom..."} This underscores the need for manufacturer sensitivity to privacy concerns.

\subsubsection{No Use Case}
3 users (1.85\%) saw no purpose in a bedroom camera, indicating that manufacturers should emphasize device utility and ensure user consultation.

\begin{figure}[h]
\centering
\includegraphics[width=0.96\linewidth]{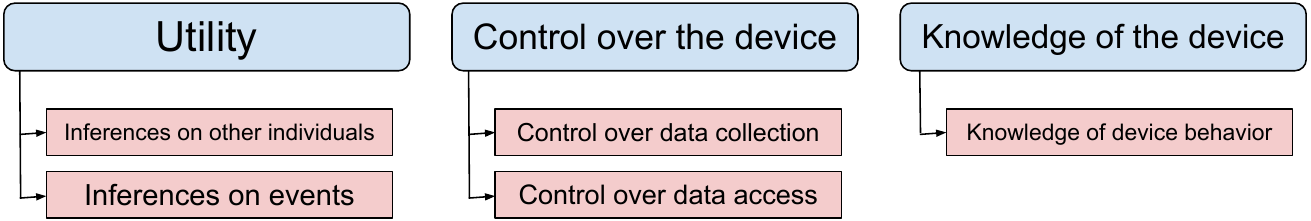}
\caption{Inductive coding of positive sentiment responses: themes (blue) and codes (red).}
\label{fig:code_pos}
\end{figure}

\subsection{Reasons For Camera Placement}

Positive responses were distilled into five codes and three themes, detailed in \figref{fig:code_pos}.

\subsubsection{Utility}
Some saw potential benefits, such as monitoring for intruders, emphasizing that perceived utility can drive adoption.

\subsubsection{Control over Device}
The majority valued control over data collection (N=19) and access (N=40). Comments included \textit{"I control what it sees so it doesn't worry me."} This supports the idea that control over a device enhances user trust.

\subsubsection{Device Knowledge}
11 users trusted the camera due to their understanding of its operation. Comments like \textit{"If I know what it is doing at all times I wouldn't mind..."} suggest that informed users are more likely to trust a device.

\section{Discussion}\label{sec:dicussion}

In this section, we discuss the key outcomes from our work, their implications, and the study's limitations.

\subsection{Key Findings}

\begin{itemize}
    \item \textbf{Data Representation vs Inference:} Perceptions of privacy are influenced not only by the interpretability of a sensor's data representations but also by possible inferences. In cases where data is non-interpretable, inferences significantly shift user privacy perceptions.
    
    \item \textbf{Physical Appearance and Familiarity:} Users' willingness to adopt a sensor is closely tied to their familiarity with its appearance. The physical form plays a pivotal role in shaping privacy perceptions, especially when the device resembles another familiar device.
    
    \item \textbf{Control:} Empowering users with control over the device design and data policies enhances their comfort and trust, making them more likely to use the device.
\end{itemize}

\subsection{Limitations of our Study}

Our study primarily relied on presenting sensor images, data, and inferences to participants. A more immersive approach could involve showing users their own data and derived inferences. Understanding how privacy perceptions change with the environment (e.g., bedroom vs. balcony) is also an avenue for future research.

\subsection{Implications for IoT/Sensing Privacy Research}

\begin{itemize}
    \item \textbf{Avoiding Assumptions:} We caution against generalizing privacy perceptions of devices like RF sensors without user-centric studies. User opinions are paramount in shaping the privacy landscape of sensors.
    
    \item \textbf{Privacy in the ML era:} The rise of machine learning necessitates a shift from evaluating only human interpretations of data to considering machine-derived inferences. The expanding capabilities of ML models, like DALL-E, mean that sensor data can carry deeper semantic meanings, underscoring the need for advanced privacy-preserving methods.
    
    \item \textbf{Transparency and Disclosure:} Besides the standard data collection information, manufacturers should be explicit about the inferences drawn from the data, potential inferences, and analogs in functionality. Clarity in disclosure, especially when focusing on inferences, is vital in privacy-sensitive contexts.
    
    \item \textbf{Empowering Users:} Prioritizing user control—be it design flexibility, data deletion rights, or opting out entirely—can significantly bolster trust and adoption rates. Ensuring user consent before data utilization is crucial.
\end{itemize}
\section{Conclusion}\label{sec:conclusion}

In this paper we conduct a user study to show that the idea that the privacy sensitivity or the privacy invasiveness of a device depends only upon the human intepretability of the data collected by that device is incorrect. Additionally, we show that a user's privacy perceptions regarding a device or a sensor depend upon a combinations of multiple different factors such as the data collected by the device, the inferences that can be made of that data, user's familiarity with the device and the amount of control that the user has over the design and data policies of the device.  This is in contrast to existing notions of privacy which assumes that human-interpretability is primary in determining the privacy-invasiveness of a sensor. We hope that in light of the key findings in this paper, manufacturers will improve their disclosure process by adding the key factors highlighted in this paper to better guide decision making when it comes to privacy.
\section{Acknowledgements}

The research reported in this paper was sponsored in part by: the NIH mHealth Center for Discovery, Optimization and Translation of Temporally-Precise Interventions (mDOT) under Award \#1P41EB028242; the National Science Foundation (NSF) under awards \#1705135 and \#1822935; and the IoBT REIGN Collaborative Research Alliance funded by the DEVCOM Army Research Laboratory under Cooperative Agreement W911NF-17-2-0196. The views and conclusions contained in this document are those of the authors and should not be interpreted as representing the official policies, either expressed or implied, of the funding agencies.
\bibliographystyle{IEEETran}
\bibliography{Sources/references.bib}




\end{document}